\documentclass[12pt]{article}
\usepackage{amssymb}
\usepackage{epsfig}
\usepackage{graphicx}

\def \bea {\begin{eqnarray}}
\def \eea {\end{eqnarray}}
\def \bea* {\begin{eqnarray*}}
\def \eea* {\end{eqnarray*}}

\def \be {\begin{equation}}
\def \ee {\end{equation}}
\def \bes {\begin{equation*}}
\def \ees {\end{equation*}}
\def \lf  {\left }
\def \rt  {\right }

\def \b  {\beta}
\def \pa  {\partial}
\def \f  {\frac}

\begin{document}

\begin{titlepage}

\begin{center}

\vskip 1.5 cm {\Large \bf All the Stationary Vacuum States of De Sitter Space}
\vskip 1 cm {Maulik Parikh\footnote{maulik.parikh@asu.edu}$^{a}$ and Prasant Samantray\footnote{prasant.samantray@asu.edu}$^{b}$}\\
{\vskip 0.75cm $^{a}$ Department of Physics and Beyond: Center for Fundamental Concepts in Science\\
Arizona State University, Tempe, Arizona 85287, USA}
{\vskip 0.75cm $^{b}$ Department of Physics\\ 
Arizona State University, Tempe, Arizona 85287, USA} 
\end{center}

\vskip  .25 cm

\begin{abstract}
\baselineskip=16pt
\noindent
We obtain all the stationary vacua of de Sitter space by classifying the inequivalent timelike isometries of the de Sitter group. Besides the static vacuum, de Sitter space also admits a family of rotating vacua, which we use to obtain Kerr-de Sitter solutions in various dimensions. By writing the metric in a coordinate system adapted to the rotating Hamiltonian, we show that empty de Sitter space admits not only an observer-dependent horizon but also an observer-dependent ergosphere.
\end{abstract}

\end{titlepage}

\section{Introduction}
The Hamiltonian, as the generator of time translations, depends of course on the definition of time. In a theory with diffeomorphism invariance, this is arbitrary. However, in suitably symmetric spacetimes, there exists a preferred class of Hamiltonians, namely those that generate timelike isometries. Timelike isometries are special because they allow quantum fields to be decomposed in terms of positive and negative energy modes in a time-independent manner. In particular, the state annihilated by a Hamiltonian that generates a timelike isometry remains empty for all time; we will refer to such states as {\em stationary vacua}.

Does every timelike isometry yield a different vacuum state? Consider Minkowski space. This has a timelike Killing vector, $\pa_T$, where $T$ is a Cartesian time coordinate. But one can also define a new time coordinate $T' = \f{T - \b X}{\sqrt{1 - \b^2}}$ and ask the following question: is the vacuum defined by the new Hamiltonian $\pa_{T'}$ different from the one defined by $\pa_T$? The answer, of course, is no, since these two Hamiltonians are related by an isometry transformation of Minkowski space, viz. a Lorentz transformation. We will call such Hamiltonians {\em group equivalent.} Isometry generators that can be mapped to each other by isometry transformations, like $\pa_T$ and $\pa_{T'}$, are part of the same equivalence class. This breaks up the isometry group into distinct conjugacy classes i.e. equivalence classes of isometrically-related generators.

In addition to group equivalence, we also have the notion of particle equivalence. Formally, two vacua defined by different choices of time are particle-inequivalent if there exists a non-trivial Bogolubov transformation (with nonvanishing $\beta_{k k'}$) between the two. The two choices of time imply different notions of positive frequency and, consequently, the vacuum state of one appears to contain particles for the other. We will refer to the corresponding Hamiltonians as {\em particle inequivalent.} 

Group equivalence implies particle equivalence. To see this, suppose two timelike isometries $H$ and $H'$ are related by a continuous isometry (i.e. they are group equivalent). But isometries leave unchanged the metric and hence also the wave operator. This means that an individual mode solution of the wave equation will, under the isometry that relates $H$ and $H'$, be mapped to another individual mode solution. In particular, a continuous isometry must map positive-frequency modes to positive-frequency modes. Hence the modes of $H$ and $H'$ are related by a trivial Bogolubov transformation; they are particle equivalent.

Although group equivalence implies particle equivalence, the converse is not true, as we shall see.
However, since group inequivalence is a necessary condition for particle inequivalence, a starting point for classifying the particle-inequivalent stationary vacuum states is to enumerate all the group-inequivalent generators of timelike isometries. This involves identifying the conjugacy classes of the isometry group. In this paper, we consider such stationary Hamiltonians for de Sitter space in various spacetime dimensions; we find a family of vacuum states, which we call $\beta$-vacua, that are generalizations of the static vacuum state.

As a warm up, let us illustrate these ideas by enumerating all the stationary vacua of Minkowski space. The most general continuous isometry of Minkowski space is generated by a linear combination of translations, boosts, and rotations:
\be
\alpha^\mu P_\mu + \beta^i K_i + \omega^{ij} J_{ij} \; .
\ee
This needs to be timelike, at least in some suitable region, for the generator to be a candidate Hamiltonian. Choosing the Hamiltonian to be $P_0$ yields the usual Poincar\'e-invariant vacuum. Alternatively, we note that the boost generator, $K_i$, squares to $X_0^2 - X_i^2$, which is timelike when restricted to the wedges $X_i^2 > X_0^2$ and is future-directed when further restricted to $X_i > 0$. This is of course the right Rindler wedge. Note that the orbit of $K_i$ starting from a point in the right Rindler wedge remains in the wedge. Choosing the Hamiltonian to be $K_i$ yields the Rindler vacuum for the right Rindler wedge, while choosing the Hamiltonian to be $-K_i$ gives the Rindler vacuum for the left Rindler wedge. Another possibility is the combination $K_i + \alpha P_0$, but this is conjugate to $K_i$, the Rindler Hamiltonian. For a less familiar example, consider the generator $K_i + \alpha P_j$ for $i \neq j$. This generates the worldline of an accelerating Rindler observer with a constant drift velocity parallel to the Rindler horizon \cite{Korsbakken, Russo}. It can be shown that this generator cannot be reduced to either $P_0$ or $K_i$ by other isometries. However, it turns out that the vacuum defined by this generator is particle-equivalent to the Rindler vacuum; they are related by a trivial Bogolubov transformation \cite{Pfautsch}. This is a case in which the Hamiltonians are group inequivalent but particle equivalent. It is straightforward to check that there are no other inequivalent isometric Hamiltonians for Minkowski space. For example, the combination $P_0 + \omega J_{12}$, which generates the worldlines of observers rotating in the $X_1-X_2$ plane with angular velocity $\omega$, becomes spacelike outside the sphere $X_1^2 + X_2^2 = 1/\omega^2$; restriction to the world-volume of the inside of the sphere fails because such a region does not admit a Cauchy surface. Or, the combination $P_0 + \alpha^i P_i$ is timelike for $\alpha^i \alpha_i < 1$ but this is obviously isometrically-equivalent to the Poincar\'e Hamiltonian via a Lorentz boost. It is easy to check that there are no other inequivalent isometries that could be used as the Hamiltonian. In summary, the only stationary vacua of Minkowski space are the Poincar\'e-invariant vacuum and the Rindler vacuum. 

Similarly, the conjugacy classes of the isometry group of anti-de Sitter space can be used as a starting point to classify the stationary vacua of anti-de Sitter space \cite{ParikhSamErik}. In this case, there is the vacuum annihilated by the generator of global time, which is the AdS counterpart of the Poincar\'e-invariant vacuum. There is also a Rindler-AdS vacuum. But, remarkably, in three spacetime dimensions, there exists a one-parameter family of $\beta$-vacua, which give a kind of rotating Rindler-AdS space \cite{ParikhSamErik,ParikhSamantray}. With this time coordinate, AdS$_3$ possesses not only an observer-dependent Rindler horizon but also an observer-dependent ergosphere.

The goal of this paper, then, is to use a similar group-theoretic analysis to enumerate the stationary vacua of de Sitter space. As in AdS$_3$, we will find that de Sitter space (in all dimensions) admits a family of $\beta$-vacua, leading to Kerr-de Sitter space, a kind of rotating de Sitter space. To be clear: the vacuum states we are looking for are generalizations of the static vacuum of de Sitter space. The Bunch-Davies state and the $\alpha-$vacua \cite{Allen} are not part of this set. They are not stationary vacua because, for example, the positive-frequency Bunch-Davies modes are eigenmodes (at early times) of the generator of conformal time, $\frac{\partial}{\partial \eta}$, but this does not generate an isometry \cite{universal}. Our strategy is to find the stationary vacua of de Sitter space by first identifying all the group-inequivalent timelike isometries of de Sitter space. Now de Sitter space in $d$ dimensions can be described by a hyperboloid embedded in $d+1$ dimensional Minkowski space:
\be
-X_0^2 + X_1^2 + ... + X_{d}^2 = 1 \; . \label{embed}
\ee
The de Sitter group is manifestly $O(1,d)$, which is of course also the Lorentz group of the higher-dimensional Minkowski space. Hence finding the group-inequivalent isometries of de Sitter space amounts to finding the conjugacy classes of the Lorentz group. 

\section{Conjugacy Classes of the Lorentz Group}

The Lorentz group has a fascinating structure. Even for the familiar Lorentz group of four-dimensional Minkowski space, there are, in fact, five types of Lorentz transformations. That is, group elements of $SO(1,3)$ fall into five distinct conjugacy classes. One conjugacy class consists of the elliptic transformations. This is the set of Lorentz transformations conjugate to the pure rotations i.e. the elliptic transformations consist of pure rotations, $J_i$, as well as all Lorentz transformations, $\Lambda J_i \Lambda^{-1}$, that can be obtained from pure rotations via Lorentz transformations. Another  conjugacy class is that of the hyperbolic transformations; these consist of the pure boosts and their conjugates, $\Lambda K_i \Lambda^{-1}$. There is also the class of parabolic transformations, whose representative elements are the so-called null rotations, generated by $J_i + K_j$ for $i \neq j$. Most interesting for our purposes are the so-called loxodromic transformations. These are Lorentz transformations generated by commuting pairs of rotations and boosts, such as $K_z + \beta J_z$. The loxodromes cannot be reduced to either pure rotations or pure boosts by Lorentz transformations since those belong to different conjugacy classes. These, then, are the four nontrivial conjugacy classes of $SO(1,3)$. (Strictly speaking, the number of conjugacy classes is continuously infinite, as each loxodrome rotation parameter $\beta$ corresponds to its own conjugacy class.) Finally, there is also the trivial conjugacy class containing the identity transformation.

There is a nice electromagnetic analog to the Lorentz group. The Lorentz generators, $M_{\mu \nu}$, which are anti-symmetric, can be thought of as the electromagnetic field strength, $F_{\mu \nu}$; the Lorentz boosts are then like the electric field with the rotations like the magnetic field. Then, just as there are five kinds of Lorentz transformations, there are five kinds of electromagnetic field configurations. To count these, recall that the two electromagnetic Lorentz-invariants are $F \wedge {}^* F \sim E^2 - B^2$ and $F \wedge F \sim E \cdot B$. Besides the trivial configuration ($\vec{E} = \vec{B} = \vec{0}$), the four types of nontrivial electromagnetic fields are therefore i) magnetic field/elliptic ($\vec{E} \cdot \vec{B} = 0, E^2 - B^2 < 0$), ii) electric field/hyperbolic ($\vec{E} \cdot \vec{B} = 0, E^2 - B^2 > 0$), iii) radiation field/parabolic ($\vec{E} \cdot \vec{B} = 0, E^2 - B^2 = 0, \vec{E}, \vec{B} \neq \vec{0}$), iv) non-null field/loxodromic ($\vec{E} \cdot \vec{B} \neq 0$). If $\vec{E} \cdot \vec{B} \neq 0$, no Lorentz transformation can transform the field into a configuration that is either a pure electric field, a pure magnetic field, or pure electromagnetic radiation, since these all have $\vec{E} \cdot \vec{B} = 0$. Correspondingly, the loxodromes of the Lorentz group are generated by linear combinations of generators that have $\vec{J} \cdot \vec{K} \neq 0$. 

Specifically, a loxodromic generator in 3+1-dimensional Minkowski space can be written as
\be
M_{01} + \beta M_{23} \; , \label{rotaboost}
\ee
in Cartesian coordinates, where $M_{ij} = -M_{ji}$ are the usual Lorentz generators. This can be extended to higher dimensions as well. In six spacetime dimensions, we can write a loxodromic generator as
\be
M_{01} + \beta_1 M_{23} + \beta_2 M_{45} \; .	\label{6Dlox}
\ee
The key property is that the Lorentz generators appearing in the linear combination of a loxodrome have no common indices and therefore commute with each other. For odd $d$, we can always form the Lorentz-invariant Casimir 
\be
\epsilon_{i_1 \ldots i_{d}} \omega^{i_1 i_2} \ldots \omega^{i_{d-1} i_d} \; , \label{Casimir}
\ee
where $\omega$ is the parameter for the most general generator $\frac{1}{2} \omega_{ij}M^{ij}$. For example, the generator (\ref{rotaboost}) has an invariant equal to $2\beta$. We shall consider the case of even $d$ later.

\section{Stationary Vacua in de Sitter Space}

So far, we have discussed the need to characterize all the inequivalent timelike isometries. But, in order for the generator of a timelike isometry to lead to a stationary vacuum, certain additional conditions have to be satisfied:
\begin{enumerate}
\item
The surfaces of constant time should be spacelike in some region.

\item
The region must admit a Cauchy surface.

\item
The integral curves of the Hamiltonian must not exit that region.

\item
The Hamiltonian should be spacelike at future and past null infinity $\cal I^{\pm}$.
\end{enumerate}

The justification for these assumptions is the following \cite{ParikhSamErik}. In de Sitter space, there is no global timelike Killing vector. Hence we can only insist that the Hamiltonian be timelike in certain regions. In fact, it may even be spacelike within our allowed region, as in an ergoregion. So the criterion we need is that, in the allowed region, surfaces of constant $t$ (where $H = i \frac{\partial}{\partial t}$) must be spacelike. Moreover, the region of interest must admit a Cauchy surface because otherwise we would not be able define quantum states. In addition, the integral curves of the Hamiltonian should not exit that region, or we would not be able to define the time-evolution of quantum states. Finally, the last condition is justified as follows. Since de Sitter space has no global timelike Killing vector, any timelike Killing vector would become spacelike outside some region. In particular, it becomes timelike at future and past null infinity. The last condition is also consistent with the holographic principle in de Sitter space \cite{dS-CFT}. The time translation generator of the boundary conformal field theory (living on $\cal I^{\pm}$) is dual to the Hamiltonian generator in the bulk de Sitter space which becomes spacelike at future and past null infinity. 

\subsection{$dS_3$}

We begin with de Sitter space in three spacetime dimensions (dS$_3$). Its (connected) isometry group is $SO(1,3)$. The simplest stationary Hamiltonian which satisfies all the necessary conditions is
\be
H = M_{01} \Rightarrow \f{\pa}{\pa t} = \lf(X^{1}\pa_{0} + X^{0}\pa_{1} \rt) \;, \label{Hnonrot} 
\ee
where we have used the standard definition of the Lorentz generators as $M^{\mu\nu} = i(X^\mu \pa^\nu - X^\nu \pa^\mu)$. This Hamiltonian leads to the static patch of de Sitter space. To see this, note that the requirement that this generator be timelike in certain regions yields
\be
|H|^2 = -X^2_{1} + X^2_{0} < 0 {\textrm~i.e.~} X^2_{1} - X^2_{0} = \xi^2_0 \label{cond-1}
\ee
Since the Hamiltonian involves only the $X^0,X^1$ coordinates, we can write
\begin{eqnarray}
X^0 &=& g(t) \nonumber \\
X^1 &=& f(t)
\end{eqnarray}
For this to be an isometry, the metric should be independent of the parameter $t$. We can therefore write
\be
-dX^2_0 + dX^2_1 = - \kappa^2 \xi^2_0 dt^2 \label{time-indep-metric}
\ee
where $\kappa$ is a constant with dimensions inverse length and $\xi_0$ depends possibly on other coordinates but not on $t$. We therefore find
\begin{eqnarray}
X^0 = \xi_0 \sinh \kappa t \nonumber \\
X^1 = \xi_0 \cosh \kappa t
\end{eqnarray}
so that the integral curves of $H$ describe Rindler trajectories in the higher-dimensional Minkowski space. The rest of the coordinates can then be parameterized to satisfy the embedding equation (\ref{embed}), giving
\be
ds^2 = -\left(1 - R^2 \right)dt^2 + \f{dR^2}{1 - R^2} + R^2 d\phi^2 \; ,
\ee
where we defined $\xi_0 = \sqrt{1-R^2}$.

However, since the symmetry group of dS$_3$ is $SO(1,3)$, we can also consider Hamiltonians which belong to the loxodromic conjugacy class of the Lorentz group in $M_4$. Such a generator can be written as
\be
 H = M_{01} + \b M_{23} \Rightarrow \f{\pa}{\pa t} = \lf(X^{1}\pa_{0} + X^{0}\pa_{1} \rt) - \b \lf(X^{2}\pa_{3} - X^{3}\pa_{2} \rt) \;, \label{Hrot}
\ee
where $\b$ is a parameter. The requirement that this generator be timelike yields
\be
|H|^2 = -X^2_{1} + X^2_{0} + \b^2 \left(X^2_{2} + X^2_{3}\right)<0 \; . \label{cond-1}
\ee
We see that $|H|^2$ becomes positive (i.e. the generator becomes spacelike) for large values of $X^0$, which is one of our requirements. Note that $H$ cannot be reduced to (\ref{Hnonrot}) by any isometry transformation. This is guaranteed by the existence of a non-zero Casimir $\epsilon_{ijkl} {\omega}^{ij}{\omega}^{kl} = 2\b$, where $\omega_{ij} = -\omega_{ji}$ are the usual parameters of the Lorentz generators in 3+1 dimensions. Therefore, the Hamiltonians (\ref{Hnonrot}) and (\ref{Hrot}) are group inequivalent. But for the vacua described by (\ref{Hnonrot}) and (\ref{Hrot}) to be inequivalent, there has to be a nonzero Bogolubov beta coefficient between the two, or in other words they have to be particle inequivalent. To calculate the Bogolubov coeficients, we follow our earlier steps to coordinatize dS$_3$ described by  (\ref{Hrot}) as  
\begin{eqnarray}
X^0 &=& \sqrt{\f{1 - r^2}{1 + \b^2}} \sinh ( t - \b \phi) \nonumber \\
X^1 &=& \sqrt{\f{1 - r^2}{1 + \b^2}} \cosh ( t - \b \phi) \nonumber\\
X^2 &=& \sqrt{\f{r^2 + \b^2}{1 + \b^2}} \cos(\phi + \b t) \nonumber\\
X^3 &=& \sqrt{\f{r^2 + \b^2}{1 + \b^2}} \sin(\phi + \b t) \; . \label{KerrdScoords}
\end{eqnarray}
The metric then reads
\be
ds^2 = -\f{(r^2 + \b^2)(1 - r^2)}{r^2}dt^2 + \f{r^2 dr^2}{(r^2 + \b^2)(1 - r^2)} + r^2 \left(d\phi + \f{\b}{r^2}dt \right)^{\! 2} \; , \label{KerrdS}
\ee
where $\phi \sim \phi + 2\pi$. This metric describes Kerr de Sitter space \cite{Park,VJ-Kerr-dS} in 2+1 dimensions, without any point defect. Note that Kerr-de Sitter space has an ergoregion where the norm of $\pa_t$ vanishes. 
The mass and angular momentum of the spacetime are
\begin{eqnarray}
M &=& \f{1 - \b^2}{8G}\\
J &=& \f{\b}{4G} \; .
\end{eqnarray}
(Following the definitions of \cite{VJ-Kerr-dS}, empty de Sitter space has nonzero mass.)

It is important to recognize that (\ref{KerrdScoords}) is just ordinary, empty de Sitter space expressed in unusual coordinates. In some sense, it is not a different spacetime, much as Rindler space is locally just Minkowski space in unusual coordinates. Indeed, by means of a diffeomorphism, $t\rightarrow t - \beta \phi$ and $\phi \rightarrow \phi + \beta t$, static de Sitter space and Kerr-de Sitter space can be mapped to each other, even globally. Nevertheless, there is an important distinction between the two spacetimes: the diffeomorphism that relates them is not an isometry. So, in particular, the vacuum states corresponding to the different time coordinates are not group-equivalent;  given a holographic dual theory to de Sitter space, the vacuum states in the dual theory would not be related by conformal transformations.

Are the Kerr-de Sitter vacua labeled by the parameter $\b$ particle-equivalent to the static vacuum? Consider a positive-frequency ($\omega > 0$) mode of the Klein-Gordon equation static coordinates:
\be
u_{n,\omega}(t, \phi, r) = e^{-i\omega t + in\phi} f_{n,\omega}(r) \; ,
\ee
where $n$ is any integer. Under the transformation $t\rightarrow t - \beta \phi$ and $\phi \rightarrow \phi + \beta t$, the spacetime becomes Kerr-de Sitter. Consider a positive energy mode ($\nu >0$) in Kerr-de Sitter coordinates:
\be
v_{l,\nu}(t', \phi', r) = e^{-i\nu t' + il\phi'} g_{l,\nu}(r) \; .
\ee
The Bogolubov beta coefficient between the two modes can be easily calculated as 
\be
\beta(n,\omega;l,\nu) =  i\Theta(-\nu - \beta l) \delta \left(\omega + \frac{\nu + \beta l}{1 + \beta^2}\right) \delta \left(n + \frac{l - \beta \nu}{1 + \beta^2}\right) \; . \label{Bogolubov} 
\ee
Note the $\Theta$ function in front: when the rotation parameter $\beta$ is zero, the $\Theta$ function vanishes. However, for nonzero $\beta$, there is a range of $l$ for which $\Theta$, and therefore also the beta coefficient, is nonzero. The $\beta$-vacua are therefore distinct from the static vacuum of de Sitter space; an observer in the static vacuum perceives a $\beta$-vacuum of Kerr-de Sitter as filled with an infinite sea of particles for each positive frequency $\omega$. 

\subsection{$dS_4$, $dS_5$} 
Next, consider dS$_4$, whose (identity-connected) isometry group is $SO(1,4)$. As always, one stationary vacuum is the static vacuum. But, taking the cue from our previous analysis, we can also consider a loxodromic Hamiltonian:
\be
H = M_{01} + \b M_{23} \Rightarrow \f{\pa}{\pa t} = \lf(X^{1}\pa_{0} + X^{0}\pa_{1} \rt) - \b \lf(X^{3}\pa_{2} - X^{2}\pa_{3} \rt) \label{candidate} 
\ee
Note that, since the embedding space has odd dimensionality, this generator does not include one of the coordinates ($X^4$ in this case). It is not at all obvious that the above generator and (\ref{Hnonrot}) belong to different conjugacy classes of $SO(1,4)$ since the Casimir $\epsilon_{abcd...}\omega^{ab}\omega^{cd}...$ does not exist in odd dimensions. In the absence of a Casimir, proving the group inequivalence of the Hamiltonians (\ref{Hnonrot}) and (\ref{candidate}) is non-trivial. Do (\ref{Hnonrot}) and (\ref{candidate}) belong to different conjugacy classes of $SO(1,4)$? To answer this question, it suffices to prove the particle inequivalence of the corresponding vacua. Consider a massive scalar field operating in the static patch of dS$_{d+1}$, noting that this spacetime is described by the Hamiltonian (\ref{Hnonrot}).

Separating variables using spherical harmonics, $Y_{l}(\Omega)$, we seek the solution for the massive Klein-Gordon equation in static coordinates as \cite{Hartnoll}
\be
\Phi(t,r,\Omega) =  \varphi (r) e^{- i \omega t} Y_{l}(\Omega) \; . \label{field}
\ee
The general solution to the radial part of the wave equation has the form
\be
\varphi = B \varphi_{\rm norm} + A \varphi_{\rm non-norm} \; , \label{eq:gensol}
\ee
where
\begin{eqnarray}
\varphi_{\rm norm} & = & \left(1- \frac{r^2}{\ell^2}\right)^{-i\omega/2} \left(\frac{r}{\ell}\right)^l  {}_2F_1\left(a + h_-, a + h_+ ; \frac{d}{2} + l; \frac{r^2}{\ell^2}\right) \\
\varphi_{\rm non-norm} & = & \left(1- \frac{r^2}{\ell^2}\right)^{-i\omega/2}\left(\frac{r}{\ell}\right)^{2-d-l} {}_2F_1\left(b + h_-, b + h_+ ; \frac{4-d}{2} - l; \frac{r^2}{\ell^2}\right) \; .  \label{dSmodes} 
\end{eqnarray}
Here $a = (l-i\ell \omega)/2$, $b=(2-d-l-i\ell \omega)/2$ and the weights are
\be\label{eq:weights}
h_\pm = \frac{d}{4} \pm \frac{x}{2} \; ,
\ee
where
\be\label{eq:mm22}
\ell^2 m^2 = \frac{d^2}{4} - x^2 \; .
\ee
Based on the falloff behavior near the origin, we observe that $\varphi_{\rm norm}$ is normalizable and $\varphi_{\rm non-norm}$ is non-normalizable. Expanding the hypergeometric functions in the solutions (\ref{dSmodes}) near the horizon, as $r \to \ell$, one finds the two behaviors: $\varphi \sim \left(1-r^2/\ell^2 \right)^{\pm i\ell \omega/2}$. These are again a superposition of ingoing and outgoing plane waves if one defines a tortoise coordinate. This means that $\omega$ is independent of $l$. The spacetime described by the Hamiltonian (\ref{Hrot}) is related to the spacetime of the usual static patch by the simple transformation $\phi \rightarrow \phi + \b t$, where $\phi$ is the azimuthal angle in $\Omega$. In other words, for $\omega>0$ and $n \in \mathbb{Z}$ where $-l \leq n \leq +l$, we can have in (\ref{field})
\be
 e^{-i \omega t}e^{i n \phi} \rightarrow e^{-i(\omega - \b n)t}e^{i n \phi} \; .
\ee
This, coupled with the fact that $\omega$ is not constrained by $l$ implies that a positive energy mode in the vacuum described by the Hamiltonian (\ref{Hnonrot}) is not necessarily a positive energy mode in the vacuum described by the Hamiltonian (\ref{Hrot}), i.e. $\omega - \b n <0$ for certain values of $\omega$ and $n$. Therefore the vacua are particle inequivalent, as there exists a nonzero Bogolubov coefficient ${\cal \beta}$ between the two spacetimes. This is a general argument and holds for all spacetime dimensions. This result also ensures that the corresponding Hamiltonians, (\ref{Hnonrot}) and (\ref{candidate}), belong to different conjugacy classes of $SO(1,4)$ and are therefore group inequivalent.

A suitable coordinatization which describes this rotating vacuum is
\begin{eqnarray}
X^0 &=& \sqrt{1 - r^2} \sinh \left(t -\b \phi \right)\nonumber \\
X^1 &=& \sqrt{1 - r^2} \cosh \left(t - \b\phi \right) \nonumber \\
X^2 &=& r \sin \theta \cos \left(\phi - \b t\right) \nonumber\\
X^3 &=& r \sin \theta \sin \left(\phi - \b t\right)\nonumber\\
X^4 &=& r \cos \theta \label{HigherdimKerrds}
\end{eqnarray}
The corresponding metric is
\begin{eqnarray}
ds^2 & = & -\left (1 - r^2 - r^2 \b^2 \sin^2 \theta \right ) dt^2 + \f{dr^2}{1 - r^2} + \left( r^2\sin^2 \theta - \b^2 \left( 1 - r^2\right) \right) d\phi^2 \nonumber\\
& & + 2 \b  \left (1 - r^2 \left(1 + \sin^2\theta\right) \right) dt \, d\phi\label{HigherdimKerrdSSolution}
\end{eqnarray}
The horizon is at $r=1$ and the ergosphere is given by the surface $r^{-2} = 1 + \b^2 \sin^2\theta$. This new solution is essentially a four dimensional analogue of Kerr-de Sitter solution in three spacetime dimensions.

Similar rotating vacua in five-dimensional de Sitter space can be constructed by considering a candidate Hamiltonian of the form  
\begin{eqnarray}
H &=& M_{01} + \b_1 M_{23} +\b_2 M_{45} \nonumber \\
\Rightarrow \f{\pa}{\pa t}  &=& \lf(X_{1}\pa_{0} + X_{0}\pa_{1} \rt) - \b_1 \lf(X_{2}\pa_{3} - X_{3}\pa_{2} \rt) - \b_2 \lf(X_{4}\pa_{5} - X_{5}\pa_{4} \rt) \; .
\end{eqnarray} 
However, in this case (analogous to dS$_3$), the existence of a non-zero Casimir $\epsilon_{abcdef}~\omega^{ab}\omega^{cd}\omega^{ef}$ guarantees that $H$ is group-inequivalent to the static Hamiltonian. This is in fact the situation for all the odd-dimensional de Sitter spaces. In principle, one can also construct higher dimensional rotating de Sitter spaces using similar loxodromic Hamiltonians \cite{RotDavies}. These solutions are analogous to the topological black holes in anti-de Sitter space \cite{Emparan,BTZ,BTZH}, even though they are not black hole solutions.

\bigskip
\noindent
{\bf Acknowledgments}

\noindent
We thank Vijay Balasubramanian and Paul Davies for helpful discussions. M. P. is supported in part by DOE grant DE-FG02-09ER41624.
         

\end{document}